\documentclass[lettersize,journal]{IEEEtran}
\usepackage{amsmath,amsfonts}
\usepackage{algorithmic}
\usepackage{algorithm}
\usepackage{array}
\usepackage[caption=false,font=normalsize,labelfont=sf,textfont=sf]{subfig}
\usepackage{textcomp}
\usepackage{stfloats}
\usepackage{url}
\usepackage{verbatim}
\usepackage{graphicx}
\usepackage{cite}
\usepackage{orcidlink}
\usepackage{hyperref}

\newcommand{\edit}[1]{\textcolor{black}{#1}}

\graphicspath{{./}{figures/}}

\newcommand{\nraoCV}{\edit{NRAO}, Charlottesville, VA, USA}
\newcommand{\nraoNM}{\edit{NRAO}, Socorro, NM, USA}
\newcommand{\gbo}{\edit{GBO}, Green Bank, WV, USA}
\newcommand{\spaceX}{\edit{SpaceX}, Hawthorne, CA, USA}

\newcommand{\bnhan}{Bang D. Nhan}
\newcommand{\cdepree}{Christopher G. De Pree}
\newcommand{\abeasley}{Anthony Beasley}
\newcommand{\mwhitehead}{Mark Whitehead}
\newcommand{\kryan}{Kevin Ryan}
\newcommand{\dfaes}{Daniel Faes}
\newcommand{\tchamberlin}{Thomas Chamberlin}
\newcommand{\dpattison}{Dawn Pattison}
\newcommand{\vcatlett}{Victoria Catlett}
\newcommand{\alawson}{Aaron Lawson}
\newcommand{\dbautista}{Daniel Bautista}
\newcommand{\swasik}{Sheldon Wasik}
\newcommand{\fschinzel}{Frank Schinzel}
\newcommand{\ddueri}{Daniel Dueri}
\newcommand{\miverson}{Matt Iverson}
\newcommand{\jdonenfeld}{Jacob Donenfeld}
\newcommand{\bschepis}{Brian Schepis}
\newcommand{\dgoldstein}{David Goldstein}

\newcommand{\bnhanA}{B.Nhan}
\newcommand{\cdepreeA}{C.De Pree}
\newcommand{\abeasleyA}{A.Beasley}
\newcommand{\mwhiteheadA}{M.Whitehead}
\newcommand{\kryanA}{K.Ryan}
\newcommand{\dfaesA}{D.Faes}
\newcommand{\tchamberlinA}{T.Chamberlin}
\newcommand{\dpattisonA}{D.Pattison}
\newcommand{\vcatlettA}{V.Catlett}
\newcommand{\alawsonA}{A.Lawson}
\newcommand{\dbautistaA}{D.Bautista}
\newcommand{\swasikA}{S.Wasik}
\newcommand{\fschinzelA}{F.Schinzel}
\newcommand{\ddueriA}{D.Dueri}
\newcommand{\miversonA}{M.Iverson}
\newcommand{\jdonenfeldA}{J.Donenfeld}
\newcommand{\bschepisA}{B.Schepis}
\newcommand{\dgoldsteinA}{D.Goldstein}

\begin{document}

\title{ODS: A self-reporting system for radio telescopes to coexist with adaptive satellite constellations}

\author{\bnhan\orcidlink{0000-0001-5122-9997},
\cdepree\orcidlink{0000-0003-3115-9359},
\abeasley\orcidlink{0000-0001-5844-8359}, 
\mwhitehead\orcidlink{0009-0004-7159-9150}, 
\kryan\orcidlink{0009-0008-9227-7520},\\
\dfaes\orcidlink{0000-0001-8603-803},
\tchamberlin\orcidlink{0000-0002-4051-7448},
\dpattison\orcidlink{0009-0007-9263-395}, 
\vcatlett\orcidlink{0000-0002-4925-8403},
\alawson\orcidlink{0009-0000-0879-5125},\\
\dbautista\orcidlink{0009-0007-3897-2912},
\swasik\orcidlink{0000-0001-9213-0117},
\edit{\fschinzel\orcidlink{0000-0001-6672-128X}},
\miverson\orcidlink{0009-0005-2746-9145}, 
\jdonenfeld\orcidlink{0009-0003-7987-6215},\\
\ddueri,
\bschepis\orcidlink{0009-0008-5489-2215},
and \dgoldstein

\thanks{\bnhanA, \cdepreeA, \abeasleyA, \mwhiteheadA, \alawsonA, and \swasikA~are with \nraoCV; \kryanA, \dfaesA, \dpattisonA, and \edit{\fschinzelA}~are with \nraoNM; \tchamberlinA, \vcatlettA, and \dbautistaA~are with \gbo; \miversonA, \jdonenfeldA, \ddueriA, \bschepisA, and \dgoldsteinA~are with \spaceX}}

\maketitle


\begin{abstract}
\edit{LEO NGSO} satellite constellations bring broadband internet and cellular service to the most remote locations on the planet. Unfortunately, many of these locations also host some of the world's best optical and radio astronomy (RA) observatories. With the number of LEO satellites expected to increase \edit{exponentially} in the upcoming decade, satellite downlink is a growing concern in protected radio-quiet areas like the US NRQZ. When these satellites transmit in \edit{or adjacent to the} protected RA bands, undesired out-of-band emission can leak into these protected bands and impact scientific observations. In this paper, we present a \edit{proof-of-concept system of a self-reporting framework - the} Operational Data Sharing (ODS) - that automates communication between radio telescopes and satellite operators by publishing telescope metadata to a \edit{secure} database accessible to \edit{participating} satellite operators through a \edit{REST API to coexist within the same spectrum}. Satellite operators can use the ODS data to adapt their downlink tasking algorithms in \edit{near} real time to avoid overwhelming sensitive RA facilities, such as through the novel Telescope Boresight Avoidance (TBA) technique. \edit{Results} from recent experiments between the NRAO and the SpaceX Starlink teams demonstrate the effectiveness of the ODS and TBA in reducing downlink emission in the Karl G. Jansky Very Large Array's observations in the 1990-1995~MHz and 10.7-12.7~GHz bands. This automated ODS system is \edit{being} implemented by other RA facilities and could be utilized by other satellite operators \edit{soon}.
\end{abstract}


\section{Introduction}
\label{sec:intro}
Radio observatories have typically been built in remote locations, far from population centers and in environments with fewer sources of human-made radio frequency \edit{(RF)} transmissions. The establishment of the \edit{United States (US)} National Radio Quiet Zone (NRQZ) in West Virginia (WV), Virginia (VA), and Maryland (MD) in the late 1950s was a way to set aside a large area ($\sim$ 13,000 square miles) within which the placement of fixed terrestrial radio transmitters would be strictly coordinated to protect the Sugar Grove Research Station (SGRS) for national security work and the Green Bank Observatory (GBO) for radio astronomy \edit{service (RAS) in WV.} 

In the intervening decades, the use of radio waves for governmental and commercial applications has increased dramatically. The NRQZ has remained a sanctuary from the omnipresent use of powerful fixed radio transmitters. Nonetheless, the NRQZ is not exactly \edit{free from artificial RF emission for RAS}, but it is certainly quieter than most other areas in the continental US. Even at the GBO site, \edit{Wi-Fi and HDTV signals from transmitters located inside and outside the protected zone are easily detectable. These undesirable emissions, although within their allocated bands, are commonly referred to as radio frequency interference (RFI) by passive spectrum users, which is not necessarily consistent with the conventional definition of interference among active transmitting systems.} \edit{Nonetheless}, radio telescopes in remote locations have been naturally protected from local RFI sources because their response to signals is peaked in the direction that they are pointing (the sky) and at a minimum close to the horizon.

\edit{The NRQZ has mainly been used to limit terrestrial transmission. However, it does not provide protection from artificial satellites' RFI, especially when they} can pass close to a radio telescope\edit{'s field of view (FOV)} at any particular moment. When there were just a \edit{small number of} satellites in orbit, the problem remained manageable while undesirable. 

The rise of large \edit{Non-Geostationary Orbit (NGSO)} constellations consisting of hundreds to thousands of satellites at \edit{the Low-Earth Orbit (LEO; at typical altitudes of 400-800~km) occupying many points in the sky}, such as SpaceX's Starlink, Eutelsat's OneWeb, \edit{Amazon's Kuiper, and AST SpaceMobile}, has added a new challenge \edit{to radio telescopes for detecting} faint radio signals coming from the solar system, the Milky Way\edit{,} and beyond \cite{walker2020satcon1, boley2021satellite}. Although radio observatories have been located in remote locations over the past half century, many of these regions overlap with rural areas that the \edit{US} federal government has highlighted as needing broadband coverage \edit{\cite{prieger2013broadband}}. The global coverage of these LEO NGSO systems provides an excellent \edit{and} economical way to provide broadband internet access to rural communities. \edit{Yet the increasing number of satellites pose increased risk to sensitive telescope receivers.} 

\edit{A small fraction of these satellites will occasionally pass close to or cross the telescope's boresight, or the center of its FOV. Their downlink (DL) beams could directly illuminate where the telescope's pointing at in the sky. Such a main-beam-to-main-beam interaction can potentially saturate, damage, or even destroy sensitive telescope receiver components such as frontend (FE) low-noise amplifiers (LNAs). Emission from NGSO satellites can drive the telescope FE electronics into non-linear regime, resulting in gain compression, increased electronic noise, or generation of intermodulation products. Each of these can corrupt the clean data channels in the RAS spectrum (both protected and unprotected bands). If the FE electronics are not affected, the backend analog-to-digital converter (ADC) electronics could nevertheless be saturated, which would increase the noise floor or manifest extra out-of-band emission.} 

Although many passive RFI mitigation techniques, such as kurtosis-based RFI flagging \cite{nita2010generalized}, real-time tracking and canceling of terrestrial RFI sources \cite{hellbourg2014reference}, and machine learning classification \cite{vos2019generative}, have had some success in removing RFI from scientific data, RFI-free radio spectrum for astronomical research has become an increasingly precious and diminishing resource \cite{van2009radio,  walker2020satcon1}. \edit{Nonetheless, these other techniques do not mitigate the risk of telescope hardware damage.} It is imperative to adopt an agile spectrum access framework allowing \edit{passive and active} spectrum users to be \edit{made aware and avoid one another to} share their respective licensed and unlicensed bands efficiently. 

Recently, \edit{various} techniques have been studied to mitigate the RFI environment in the vicinity of radio telescopes. For example, \edit{the} Satellite Orbit Prediction Processor (SOPP) is a predictive algorithm that both anticipates potential incoming satellite RFI and then optimizes the telescope scheduling to observe a relatively cleaner sky region \cite{gifford2024satellite}. Another example is a reactive approach in which a metasurface material on the rim of a dish antenna electromagnetically nulls the telescope's main beam from individual satellites \cite{budhu2024design}. However, for a \edit{highly subscribed} scientific telescope like the US National Science Foundation (NSF)'s Karl G. Jansky Very Large Array (VLA) in New Mexico (NM) \edit{and} the Robert C. Byrd Green Bank Telescope (GBT) at GBO, it is impractical \edit{nor sustainable} for \edit{researchers} to be selective on what sky region and frequency band to use. A more dynamic and low-impact approach on the existing telescope operation is needed for these facilities.

In recent years, under \edit{the} coordination agreement between NSF and SpaceX, NRAO and SpaceX have conducted numerous experiments to characterize their systems to determine viable strategies to achieve spectrum coexistence\edit{. For example, some early tests involved having Starlink satellites avoid illuminating the geographical area where the VLA is located while still providing satellite internet service to Starlink User Terminals placed in the Alamo Navajo Reservation, located approximately 25~miles northeast from the telescope \cite{depree2023memo223}. Through these experiments, NRAO and SpaceX developed the Zone Avoidance (ZA) strategy as a basic protection from DL RFI for the majority of the satellite orbiting above the VLA and GBT. As an advanced protection for a subset of satellites passing close to the telescope boresight, previous experiments with the GBT} demonstrated the Starlink system's ability to adjust and \edit{disable DL} transmission when \edit{the} satellites pass close to the telescope's pointing using the Telescope Boresight Avoidance (TBA) technique \cite{nhan2024toward}. In this paper, \edit{as a proof of concept (POC),} we \edit{present the Operational Data Sharing (ODS) automated self-reporting system}, which \edit{provide near real-time telescopes information database (DB) to satellite operators} through an application programming interface (API) using the representational state transfer (REST) architectural style, thus enabling awareness for \edit{RFI mitigation}. \edit{The NRAO's ODS system is the first of its kind to allow a ground-based radio telescope to communicate with a satellite constellation system in near real time. We note that the technique developed in this study is not meant to address unintended RFI emitted below 200~MHz by electronics onboard the satellites \cite{divruno2023unintended}.}

\section{Spectrum \edit{C}oexistence \edit{S}trategies}
\label{sec:coexist_strategies}
One major challenge to \edit{RAS} in recent years has been to find ways to coexist with \edit{LEO NGSO systems} that are beaming high power signals to the Earth's surface. \edit{Some recently deployed systems (like the SpaceX Starlink) have the advantage of highly adjustable DL beams with phased array antennas, such that} their direction and power level can be controlled \edit{and steered by onboard software electronically.} Figure~\ref{fig:coexistence} illustrates \edit{two potential ways for satellite constellations to cooperatively share spectrum with RAS}.

\subsection{Zone Avoidance \edit{(ZA)}}
\label{sec:zone_avoidance}
\begin{figure*} 
  \centering
  {\includegraphics[width=1.3\columnwidth]{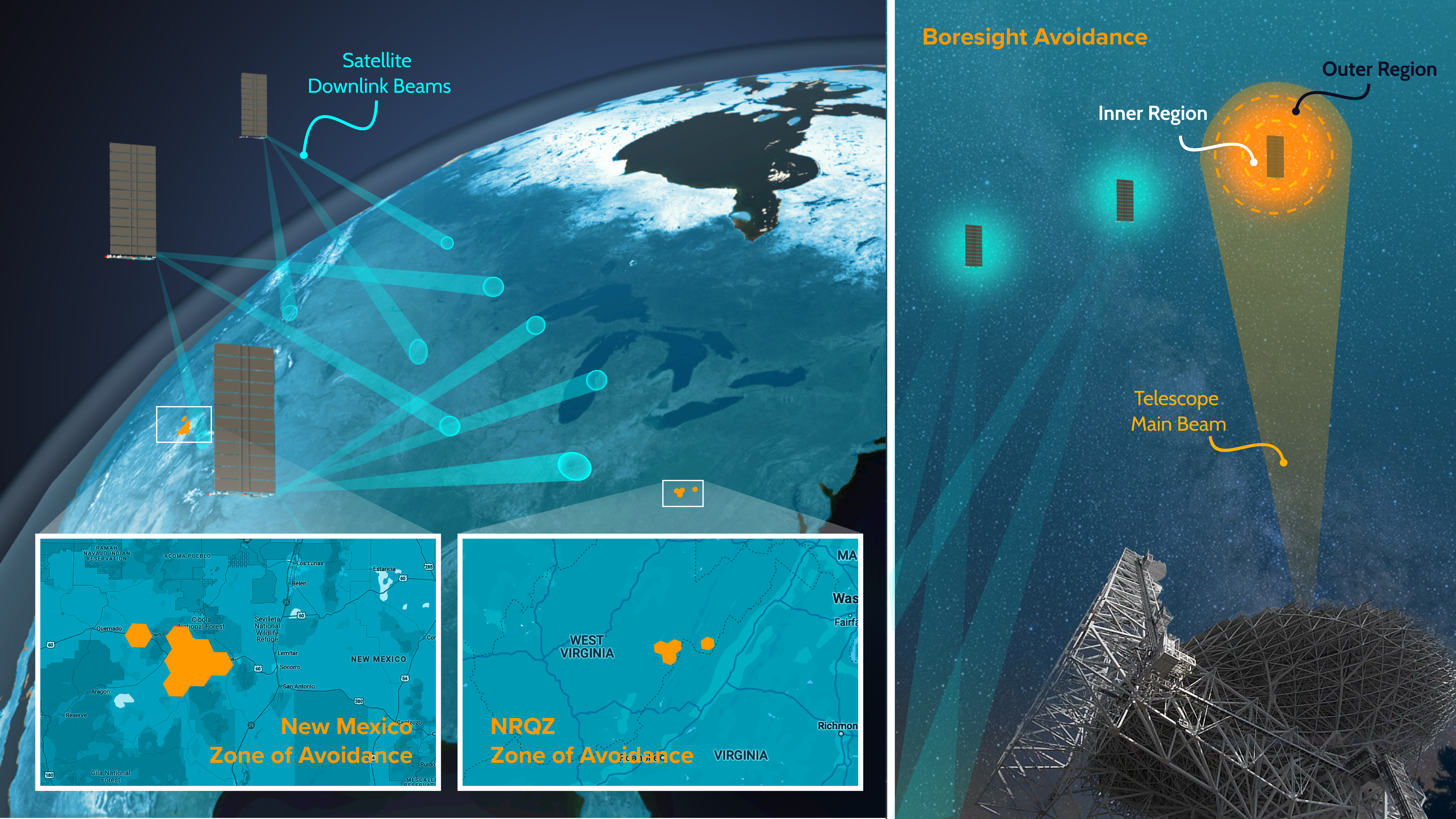}}
  \caption{A simplified illustration \edit{of} the two approaches enabling spectrum coexistence between radio telescopes and satellite operators. (\textit{Left}) Zone avoidance \edit{(ZA)} provides the \edit{basic} protection with satellites placing DL beams far away from radio telescopes residing within their service cells, highlighted by orange hexagonal hierarchical geospatial indexing regions in the lower insets, as reproduced from \href{https://www.starlink.com/map}{the Starlink Availability Map online} in February 2025. (\textit{Right}) Telescope boresight avoidance (TBA) provides \edit{an advanced} protection when a satellite passing close to the telescope main farfield beam (or boresight). The TBA consists of two modes: \edit{the} \emph{inner} boresight region (white dashed circle) where the satellite would briefly turn off the formation of phased array beams; and the \emph{outer} boresight region (orange dashed annulus) within which the satellite places its DL beams far from the telescope site.  (Credits: NRAO/ESM/Sophia Dagnello).}s
  \label{fig:coexistence}
\end{figure*}

The \edit{basic} protection approach is to have satellite constellations avoid \edit{directly illuminating the} RA sites. This partial solution requires no communication between telescopes and satellite operators, but simply a \edit{DB} of the \edit{observatory locations}. \edit{The DL beams for the Starlink's broadband internet service at 10.7-12.7~GHz have a narrow farfield beam, such that exclusion zones on the order of 10~km can be employed to protect telescopes. Based on early coordinated experiments \cite{depree2023memo154}, we determined that it is sufficient to adopt the \href{https://www.starlink.com/map}{current Starlink's US coverage map} with the highlighted} exclusion zones \edit{in place protecting US RAS sites, including} the GBO and SGRS in WV, along with the VLA in NM (see insets in Figure~\ref{fig:coexistence}). \edit{The NSF has successfully negotiated agreements with several other satellite operators to take this approach.} 

\subsection{Telescope Boresight Avoidance (TBA)}
\label{sec:boresight_avoidance}
As \edit{an advanced} protection, the TBA is a satellite tasking scheme developed by SpaceX and NRAO that allows Starlink satellites to respond to the shared ODS radio telescope information (provided at some prearranged buffer time before \edit{RA observations start}). Once the Starlink system identifies that a particular satellite trajectory will pass close to a telescope boresight operating at \edit{or near} one of its DL frequency channels, the satellite constellation takes one of three actions:
\begin{enumerate}
  \item If the satellite passes outside of an agreed ``outer boresight'' region (defined as an angular separation \edit{between the satellite and the boresight positions}), it takes no action.
  \item If the satellite crosses into the ``outer boresight'' region, it will task its beams far from the radio telescope (typically 180~km) so that the telescope is only illuminated by the DL beam's sidelobes.
  \item  If the satellite further crosses into an agreed upon ``inner boresight'' region, it will momentarily disable beam forming \edit{and DL transmission.}
\end{enumerate}
 \edit{The outer and inner TBA angular cutoffs depend on the telescope's observing receiver band and reflector size}, both of which affect its beam size, \edit{which corresponds to the FOV and duration of detecting the satellite's strong DL signal}. 
 
\begin{figure} 
  \centering
  {\includegraphics[width=1\columnwidth]{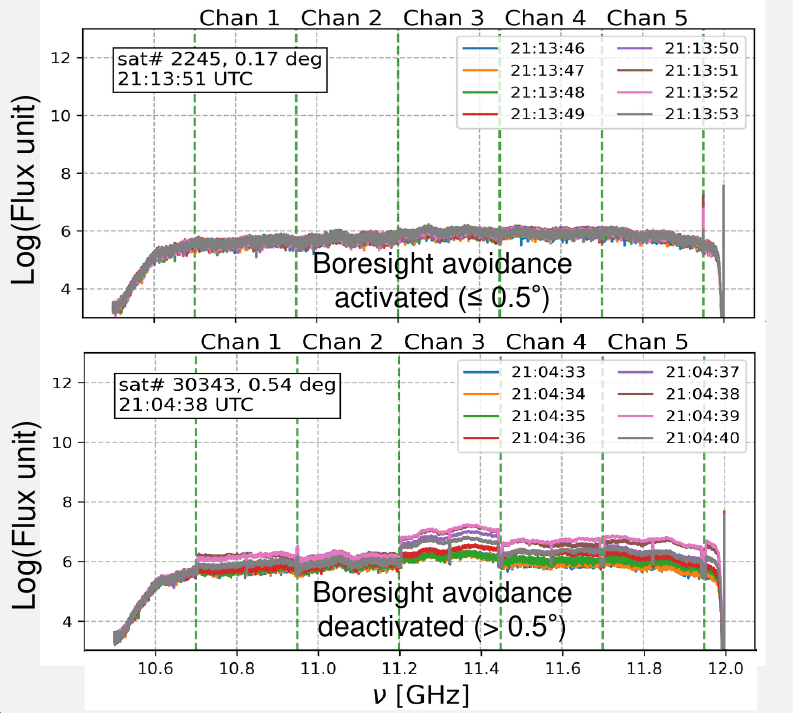}}
  \caption{Comparison between spectra measured between 10.6-12.0~GHz during two different Starlink passages \edit{from} previous coordinated GBT tests: with (boresight angular separation of $0.17^{\circ}$, top panel) and without ($0.54^{\circ}$, bottom panel) TBA activated, using an inner boresight cutoff of $0.5^{\circ}$. When the Starlink disabled its transmission (in DL channels between 10.7-12.7~GHz), the observed spectrum appears to be nominal with much less broadband RFI. This figure is reproduced from \cite{nhan2024toward}.}
  \label{fig:gbt_tba_tests}
\end{figure}

Fortunately, due to \edit{the large diameter of modern single-dish telescopes, their} main beam sizes at the Starlink's \edit{10.7-12.7~GHz internet DL bands} are \edit{relatively small thus reducing the interaction time with close-passing satellite to a few seconds. However, this is not the case for newer the Direct-to-Cell (DTC), or Supplementary Coverage from Space (SCS), service the newer Starlink satellites are providing to T-Mobile cellular devices at 1990-1995~MHz. Due to the lower frequency thus larger beam size, and the required stronger signal level for communicating to the smaller cellular devices, the telescopes can be exposed to the DTC DL signal for several minutes.} \edit{As an earlier POC experiment, Figure~\ref{fig:gbt_tba_tests} illustrates} the Starlink's capability in implementing the TBA \edit{when conducting coordinated testing with the GBT,} with a preliminary inner angular cutoff of $0.5^{\circ}$ from the telescope's boresight at 10.7-12.7~GHz\cite{nhan2024toward}. 

\subsection{\edit{A} Need for Automated \edit{Near} Real-Time Communication}
Early experiments performed at the GBT and VLA highlighted an important variable in the work to reduce strong \edit{RFI}: \edit{\textit{the relative angular separation between a satellite and a particular telescope's pointing}}. Early \edit{NRAO-SpaceX coordinated tests} showed clearly that even satellites that were serving regions far from the telescope could inject large signals into the \edit{telescope receiver} if \edit{they} passed close to the telescope\edit{'s pointing} \cite{depree2023memo154}. These \edit{studies confirms} the importance of \edit{providing the situational awareness to the satellite operators for a radio telescope's current operational information}, namely, \edit{when and} where it is pointing in the sky \edit{at what frequency band}. For example, \edit{a} constellation could adapt its \edit{DL} for an individual satellite \edit{of interest to better mitigate} its \edit{localized impact on a particular telescope} than \edit{attempting a} constellation-wide mitigation. \edit{In return, instead of requiring a clear spectrum at all times, RAS could benefit from a cleaner spectrum when it is needed during science observations outside the protected RA bands that may overlap with the DL channels.}

\section{How \edit{ODS W}orks}
\label{sec:ods_system}
\edit{ODS} is NRAO's attempt to facilitate \edit{near} real-time communication of telescope status to satellite operators. The testing of this system has been carried out with SpaceX \edit{so far, thus} for the remainder of this paper, we will be referring to the system that is currently in operation at \edit{the VLA}. We emphasize that the functionality of the \edit{ODS} at the VLA requires no special actions on the part of observers, who submit their observing schedule (Scheduling Blocks) as they always have. The NRAO systems then process these data without any \edit{intervention from observers}. The \edit{ODS} is also capable of disabling status reporting (e.g., for \edit{proprietary} reasons) for \edit {certain} observations by a particular \edit{observer}upon request. \edit{In} return, these observations \edit{will not be informed and protected by TBA}. By design, the ODS framework can be adopted by any satellite operators and radio observatories as long as they \edit{use} the same data and \edit{API} standards.

\subsection{ODS Software Architecture}
\label{sec:ods_software_framework}
\begin{figure} 
  \centering
  {\includegraphics[width=.95\columnwidth]{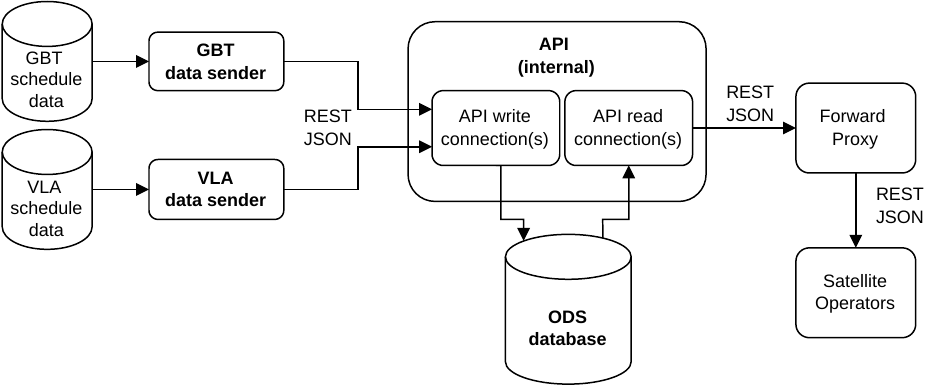}}
  \caption{A high level block diagram of the current ODS prototype \edit{shows how the telescope schedule JSON data are delivered to the ODS DB from the VLA and GBT systems, and subsequently to the satellite operators}. More details of the ODS JSON and API standards are available at \href{https://obs.vla.nrao.edu/ods/}{the ODS homepage}.}
  \label{fig:ods_framework_diagram}
\end{figure}

\edit{The \href{https://obs.vla.nrao.edu/ods/}{ODS API}, using the OpenAPI~v3.1.0 specification, acts as a wrapper for writing and reading from the DB. It can be created in multiple languages, such as JavaScript or Python. The system consists of a Data Sender program at each NRAO/GBO telescope, which first validates the telescope's near-term scheduled operational data against the predefined JSON format and requirements before \edit{passing them through the ODS API, which subsequently writes data to the ODS \edit{DB} for the satellite operator clients} at \edit{some buffer ahead of time (typically within half an hour)} before the observation starts.} \edit{This buffer time is needed for the satellite network to uplink and propagate the queried ODS data to the entire constellation.} The satellite operator(s) can constantly access and monitor the ODS \edit{DB} during the coordination to adapt the tasking of their constellation(s) in \edit{near} real-time when an \edit{ODS frequency range overlaps with their DL channels} is present. The ODS data is \edit{stored} in a dedicated DB designed for \edit{information exchange} solely between partners \edit{within a coordination agreement (e.g.,} NRAO and SpaceX). A schematic of the ODS system is shown in Figure~\ref{fig:ods_framework_diagram}.

\subsection{ODS \edit{A}doption for NRAO/GBO \edit{F}acilities}
\label{sec:ods_tests}
The \edit{ODS} is currently operational at the \edit{VLA}. The NRAO ODS system at the VLA reports upcoming observations to the \edit{API}, and SpaceX uses this information to schedule \edit{ the Starlink network to invoke TBA for both their DTC band (1990-1995~MHz) and X-band internet (10.7-12.7~GHz)}. \edit{NRAO is currently testing an implementation of the ODS system at the GBT. Additionally, new efforts are underway to incorporate the ten dedicated the Very Long Baseline Array (VLBA) sites across the US in ODS.}
\begin{figure} [!]
  \centering
  \includegraphics[width=1.1\columnwidth]  {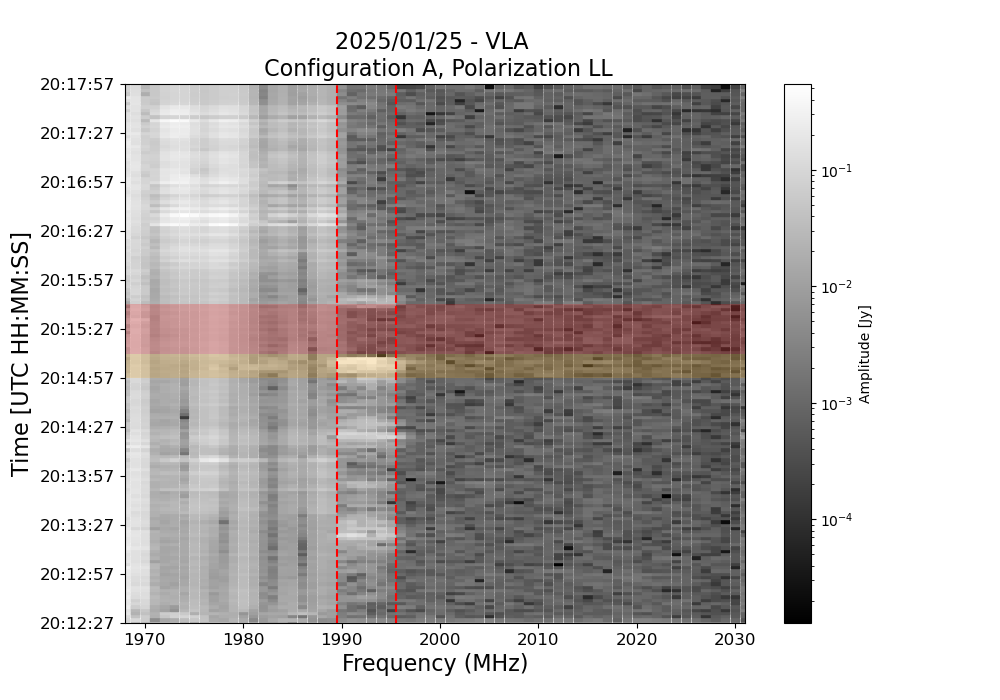}  
  \includegraphics[width=1\columnwidth]{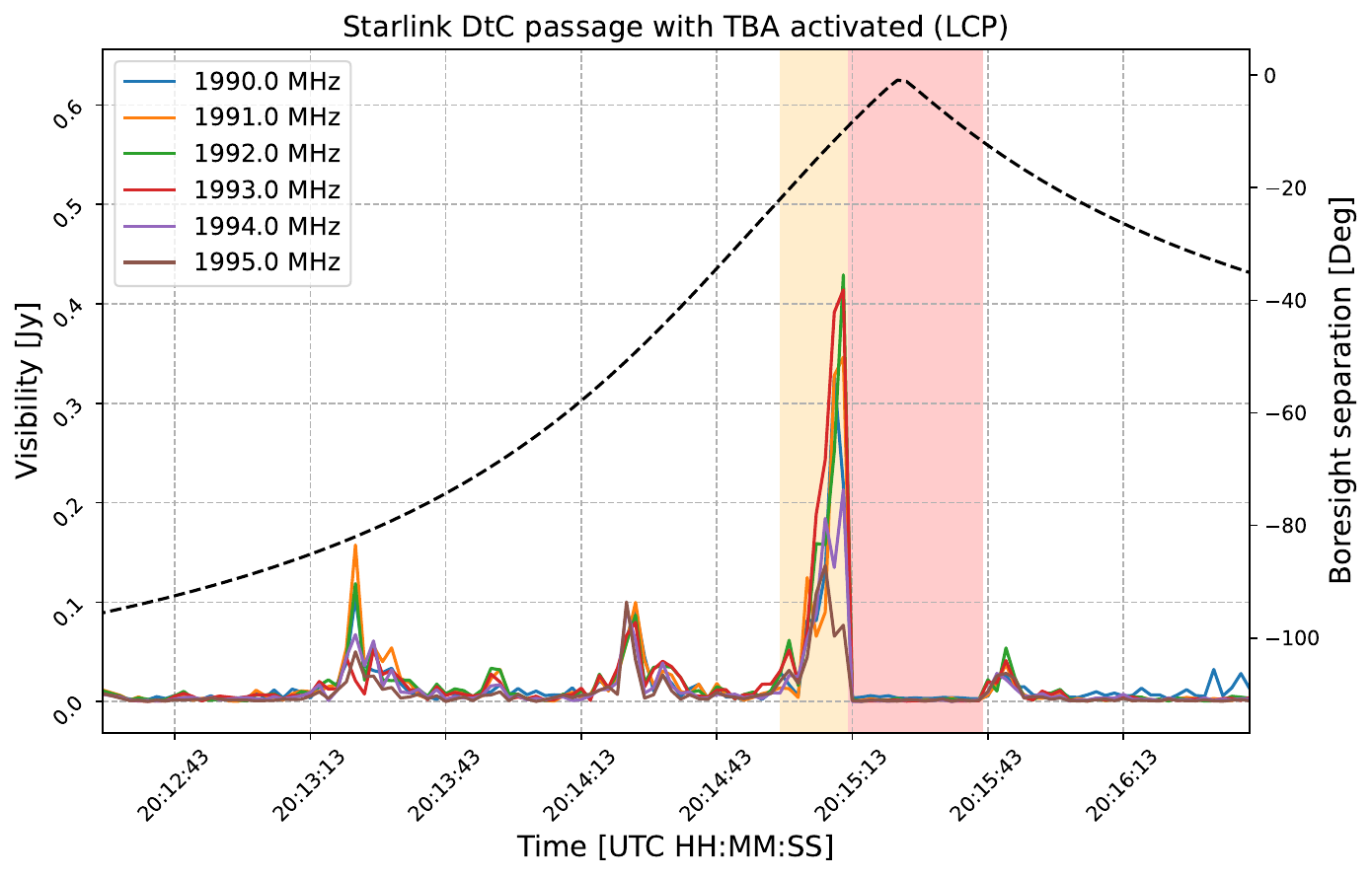}
  \caption{(\textit{Top}) Waterfall plot for the cross-correlated visibility data in left-handed circular polarization (LCP) at one of the VLA baseline pairs showing the Starlink passage with TBA activated for the D\edit{T}C band (red dashed lines) with two shaded TBA regions (using the \edit{TBA} log file provided by SpaceX), outer (yellow) and inner (red) angular cutoff regions. (\textit{Bottom}) Similarly, the profile plot for the same VLA spectra is shown as a function of time for the D\edit{T}C channels of the same Starlink passage with TBA activated with \edit{the same} shaded regions. \edit{Evidently, the angular separation between the satellite and the boresight} (black dashed curve, right $y$-axis), computed using \edit{public} TLE data, shows the expected behavior of the TBA modes.}
\label{fig:ods_tba_dtc_plots_vla_jan2025} 
\end{figure}

\subsection{ODS \edit{TBA V}erification \edit{at VLA}}
\label{sec:ods_verification}
To develop a closed-loop system, SpaceX has been providing stripped-down satellite tasking logs since mid-August of 2024. The \edit{TBA} log files only contain information for satellite passages which have activated TBA, including the timestamps, satellite ID, along with three TBA parameters: inner/outer mode, angular cutoff values, and avoided DL bands. These \edit{logs} are invaluable for the NRAO team to evaluate the effectiveness of \edit{the ODS's TBA during science observations}. \edit{To illustrate}, Figure~\ref{fig:ods_tba_dtc_plots_vla_jan2025} shows the waterfall and spectrum plots from data randomly selected in one of the VLA science observations in Jan~2025. The activation of TBA is apparent at the \edit{DTC} 1990-1995~MHz band, for this particular Starlink passage. The TBA activation times are consistent with the simulated boresight angular separation between the satellite and the telescope boresight, computed using public Starlink two-line element (TLE) data, using the Vincenty's formula implemented in \edit{\href{https://docs.astropy.org/en/stable/api/astropy.coordinates.angular_separation.html}{Astropy}}.

\begin{figure*} [!t]
  \centering
  \includegraphics[width=1.5\columnwidth]{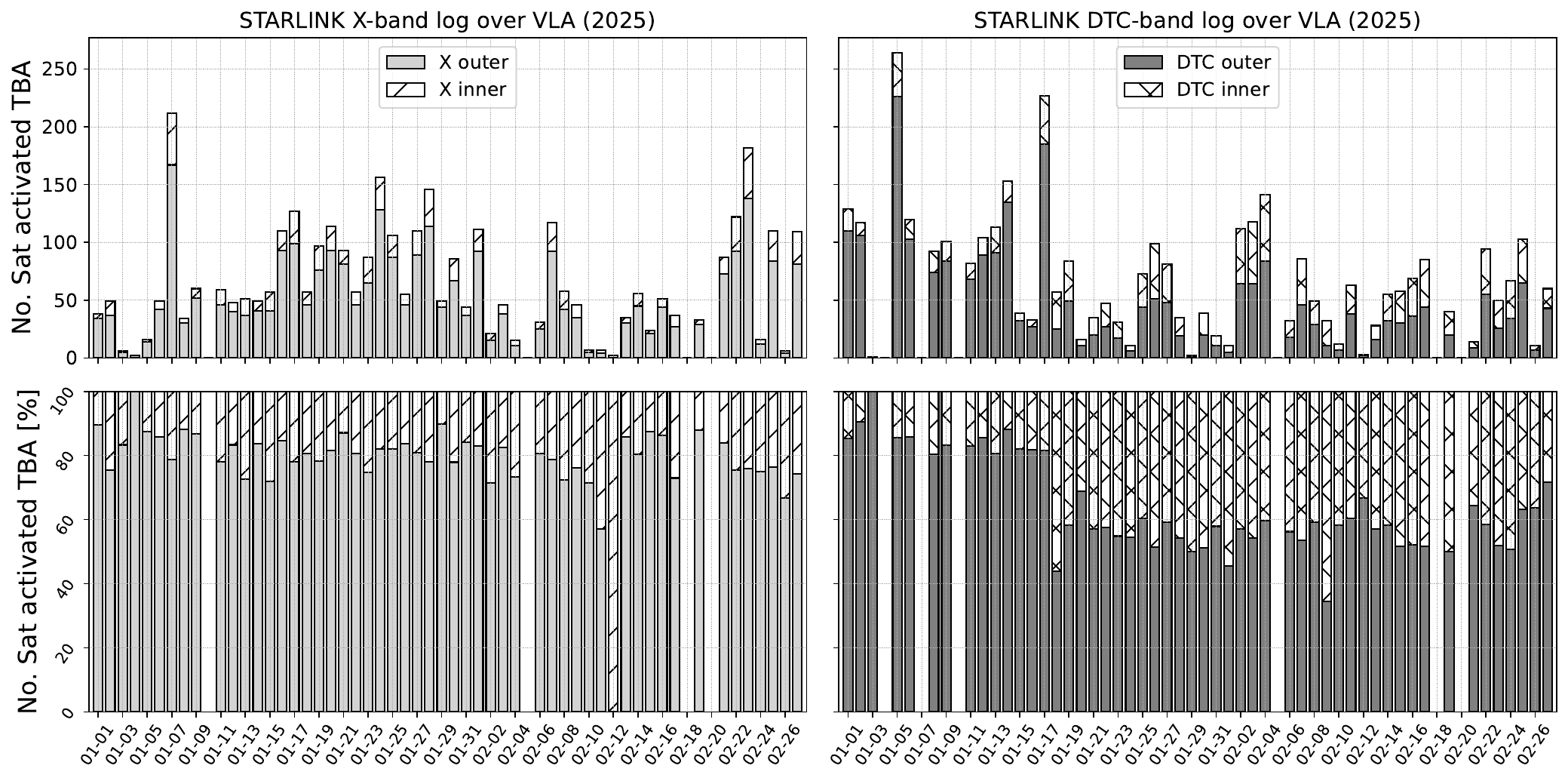}
  \caption{Statistics of Starlink satellites activated the boresight avoidance with the provided ODS log data of the VLA in January and February of 2025. (\textit{Left panels}) The number of satellites passages and relative percentage for the inner and outer boresight avoidance modes with X-band DL. (\textit{Right panels}) Similar statistics exist for the satellites with the D\edit{T}C DL. \edit{This illustrates that only a minority of the satellite passages requires to disable their downlink within the inner angular cut off region and ensuring minimal impact on the satellite network's service capacity.} Noting that there are a few days when some satellite passages were not close enough to the telescope pointings to trigger one of the TBA modes thus the values are null.}
  \label{fig:SatLog}
\end{figure*}

\begin{figure} 
  \centering
  \includegraphics[width=1\columnwidth]  {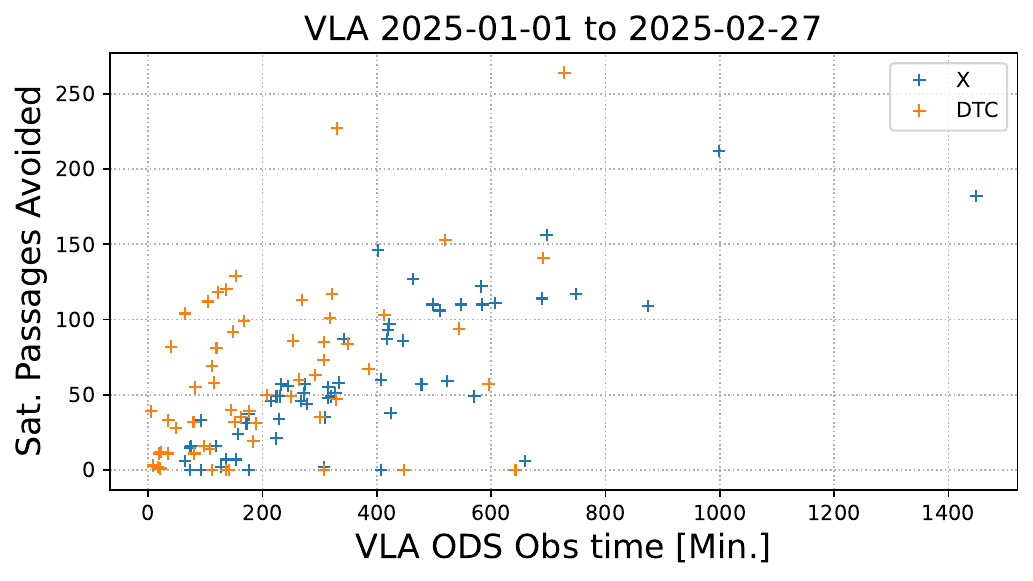}
  \caption{Comparison of the daily total number of satellite passages activated the TBA and the daily total observing time on the VLA using receivers corresponding to the D\edit{T}C at 1990-1995~MHz (orange) and X-band at 10.7-12.7~GHz (blue) range in the first 1.5 months of 2025. The approximately linear correlation of the data indicates that the Starlink satellites successfully activate their avoidance using the provided ODS data. Noting that the D\edit{T}C values are clustering to the left of the X-band values because these passages last longer per satellite due to the larger DL beam footprint at lower frequencies.}
  \label{fig:SatODSRxLog}
\end{figure}

As a result of this \edit{informed} awareness, Starlink satellites that cross into the inner boresight only need to quiet their systems within a few tens of seconds, depending on the beam size at different DL frequencies. Currently at the VLA, the inner and outer cutoffs are set at different values for the 10.7-12.7~GHz (coincides with VLA's X and Ku bands) and the 1990-1995~MHz (VLA's L and S bands) DL bands due to the different \edit{receiver beam sizes}. Figure~\ref{fig:SatLog} shows statistics for TBA at the VLA in the first 1.5~months of 2025 for both Starlink's X and D\edit{T}C bands. Note that in this Figure, there is significant daily variations in the number of avoidance events, \edit{depending} on the fraction of \edit{observation done in the respective VLA bands} (i.e., L, S, X, or Ku) that day. Similarly, Figure~\ref{fig:SatODSRxLog} shows the correlation between the total Starlink passages with TBA activated and the total observation times (in minutes) conducted by the VLA receivers at the corresponding receivers overlapping the Starlink's X and D\edit{T}C bands. The approximately linear correlation of these data indicates that the Starlink systems are successfully activating \edit{TBA} using the provided ODS data.

\section{Future Growth and Challenges}
While the NRAO ODS system has initially been prototyped to communicate activities \edit{of VLA and GBT} specifically to SpaceX, the system \edit{is designed} to be scalable. The eventual goal is for \edit{other radio} observatories to communicate their operational data to a REST API that would then be readable by different satellite constellations. \edit{With a fleet of more than 7,000 satellites, as of June 2025, t}he Starlink network is redundant and flexible \edit{enough,} in terms of beam placement\edit{, to enable its} ability to quiet \edit{onboard} electronics for short periods of time \edit{and to have negligible impact on the network's service coverage}. 

In fact, \edit{several other US and international} radio observatories have independently implemented our ODS framework based on our published API requirements and JSON format. For example, the Hat Creek Radio Observatory (\edit{\href{https://www.seti.org/hcro/ods}{HCRO}}), \edit{the \href{https://www.haystack.mit.edu/ods/}{MIT Haystack Observatory}}, and the Commonwealth Scientific and Industrial Research Organisation's (CSIRO) Australia Telescope National Facility (ATNF) are currently running ODS with TBA activated by SpaceX. \edit{The \href{https://www.narrabri.atnf.csiro.au/ods/index2.html}{CSIRO's ODS} system currently is reporting their telescope facilities including Australia Telescope Compact Array (ATCA), Parkes Observatory, Mopra Observatory, Australian Square Kilometre Array Pathfinder (ASKAP).} \edit{Meanwhile, the Owens Valley Radio Observatory (OVRO) has set-up} ODS servers,  with TBA testing underway. Currently, each observatory maintains its own ODS \edit{DB} for its telescopes. Ideally, future functionality could be simplified with the construction of a single \edit{DB} for all radio telescopes. \edit{Meanwhile NRAO will continue to develop its ODS API, improve its security, and share these developments with other observatories using the same framework.}

\edit{Currently, there is no formal international policy or regulation requiring satellite operators to adopt coexistence techniques like ZA and TBA with RAS. Not all NGSO systems may be capable of adopting these dynamic avoidance techniques. Nevertheless, NGSO operators operating in the US are required to coordinate with radio observatories under current US Federal Communications Commission (FCC) licensing requirements, although some of these coordinations have only been done on a best efforts basis. One of the goals for ODS is to establish a robust and secure information sharing standard to lower the barrier of entry for other RAS and NGSO operators.}

\section{Conclusions}
Based on early \edit{coordination} with SpaceX, NRAO has built a \edit{POC ODS} system that allows radio telescopes to observe in the \edit{DL} bands licensed to \edit{NGSO} operators that deploy \edit{TBA}. The sharing of telescope \edit{information is continuous}, but the engagement of TBA \edit{events} are only triggered when the \edit{telescope} in question is (1) observing in or near the DL bands and (2) pointed in the direction of a particular satellite trajectory. The successful integration of the ODS and TBA \edit{applied to the SpaceX DL bands of 1990-1995~MHz (DTC) and 10.7-12.7~GHz (broadband internet) for VLA observations has been verified} through analysis of SpaceX \edit{tasking} logs and RFI \edit{in the VLA spectra}. Nonetheless, the performance of the ODS and TBA \edit{systems} will require periodic monitoring. Building on this success, the NRAO will proceed with the deployment of ODS systems at the GBT and VLBA. It is imperative for NRAO to maintain direct communication with SpaceX and other satellites operators, along with other radio observatories for any future updates and standardization of the ODS system. Importantly, since the antenna feed performance varies at different frequencies from telescope to telescope, each ODS adopter will need to conduct independent tests with different satellite operators to determine suitable TBA parameters for their own instruments. 

\section*{Acknowledgments}

The \edit{NRAO and GBO are facilities} of the \edit{U.S. NSF} operated under cooperative agreement by Associated Universities, Inc. This work is supported by the NSF's SII-NRDZ (AST-2232159) and SWIFT-SAT (AST-2332422) grants. The authors acknowledge the contributions of many individuals who have made these experiments possible. At NRAO and GBO: W.Armentrout,~R.Arnold,~P.Brandt,~W.Brisken, P.Demorest, D.Frayer, J.Frothingham, M.Gardiner, Z.Graham, B.Gregory, J.Jackson, L.Jensen, J.Kern, R.Lynch, R.Minchin, T.Minter, R.Moeser, G.Monk, B.Moore, K.O'Neil, V.Parekh, Y.Pihlstrom, A.Remijan, U.Rao, J.Robnett, D.Rose, P.Salas, R.Selina, D.Schafer, F.Schwab, N.Sizemore, A.Sowinski, B.Svoboda, R.Taggart, C.Tounzen, C.Ubach, M.Wainright; and at SpaceX: M.Albulet, D.Goldman, D.Knox, T.Liang, J.McMichael, M.Nicolls, Ka.Omar, D.Partridge, \edit{and C.Zinsli}. The authors would also like to acknowledge \edit{collaborators who adopted and tested} the ODS independently for their \edit{facilities}:  D.DeBoer (UC Berkeley), P.Erickson (MIT Haystack), W.Farah (SETI), K.Gifford (CU Boulder), G.Hellbourg (Caltech), B.Indermuehle (CSIRO), A.Pollak (SETI).

\bibliographystyle{IEEEtran}
\bibliography{IEEEabrv,reference_ods} 

\section*{Biographies}
\vspace{-33pt}
\begin{IEEEbiographynophoto}{\bnhan} (\href{mailto:bnhan@nrao.edu}{bnhan@nrao.edu}) received the Ph.D. degree in Astrophysics from the University of Colorado Boulder in 2018. He is an Assistant Scientist in the Electromagnetic Spectrum Management (ESM) department at the NRAO. He is the Project Scientist for the ODS system and has been working on coordinating tests between NRAO and SpaceX. 
\end{IEEEbiographynophoto}
\vspace{-33pt}
\begin{IEEEbiographynophoto}{\cdepree} received the Ph.D. degree in Physics from the University of North Carolina at Chapel Hill in 1996. He is the Assistant
Director for ESM and the Project Director of the National Radio Dynamic Zone (NRDZ) project at NRAO. 
\end{IEEEbiographynophoto}

\vspace{-33pt}
\begin{IEEEbiographynophoto}{\abeasley} received the Ph.D. degree in Astrophysics from the University of Sydney in 1990. He is the Director of NRAO and the Project Director of the Next Generation Very Large Array (ngVLA). 
\end{IEEEbiographynophoto}

\vspace{-33pt}
\begin{IEEEbiographynophoto}{\mwhitehead} received the M.S. degree in Applied Physics from the Appalachian State University. He is  a Software Architect in the Data Management and Software (DMS) division at NRAO. He is serving as the Architect Owner of the ODS project.
\end{IEEEbiographynophoto}

\vspace{-33pt}
\begin{IEEEbiographynophoto}{\kryan} received the B.A. degree in Computer Science from the Thomas Edison State University in 1991. He has recently retired and was a Software Engineer in the New Mexico Systems (NMS) group of the Software Development Division (SDD) at NRAO. He led the development effort of the ODS API server.
\end{IEEEbiographynophoto}

\vspace{-33pt}
\begin{IEEEbiographynophoto}{\dfaes} received the Ph.D. degree in Astronomy and Astrophysics from the Universit\'{e} C\^{o}te d'Azur in 2015. He is a Software Engineer in the NMS group at NRAO. He leads efforts in developing the VLA and VLBA Data Sender modules \edit{and maintaining the original API for the ODS system}.
\end{IEEEbiographynophoto}

\vspace{-33pt}
\begin{IEEEbiographynophoto}{\tchamberlin} received the B.S. degree in Computer Science from the Georgia State University in 2014. He is a Software Engineer at the GBO. He leads efforts in developing the GBT Data Sender module and the API for the ODS system. 
\end{IEEEbiographynophoto}

\vspace{-33pt}
\begin{IEEEbiographynophoto}{\dpattison} received a B.S. in Engineering Science and Mechanics from the Virginia Polytechnic Institute and State University in 2012. She is a Software Engineer in the Scientific Support and Archive (SSA) group at NRAO. She leads the efforts in upgrading the ODS API to the FastAPI framework.
\end{IEEEbiographynophoto}

\vspace{-33pt}
\begin{IEEEbiographynophoto}{\vcatlett} received the B.S. degree in Physics and Mathematics from the University of Texas at Dallas in 2022. They were a Software Engineer at GBO. They helped to prototype the GBT Data Sender for the ODS system. 
\end{IEEEbiographynophoto}

\vspace{-33pt}
\begin{IEEEbiographynophoto}{\alawson} received the B.A. degree in Physics from the University of Virginia in 2015. He is a Scientific Data Analyst in the ESM group at NRAO. He contributes in planning and analyzing coordinated VLA observation tests with SpaceX.
\end{IEEEbiographynophoto}

\vspace{-33pt}
\begin{IEEEbiographynophoto}{\dbautista} received the B.A. degree in Physics and Astrophysics from the University of California Berkeley in 2021. He is a Scientific Data Analyst in the ESM group at GBO. He contributes in planning and analyzing coordinated GBT observation tests with SpaceX.
\end{IEEEbiographynophoto}

\vspace{-33pt}
\begin{IEEEbiographynophoto}{\swasik} received the B.S. degree in Astrophysics from the Michigan State University in 2020. He is a Zone Regulatory Service Coordinator for the \edit{NRQZ} and the Puerto Rico Coordination Zone (PRCZ) in the ESM group at NRAO. He contributes in analyzing coordinated VLA and GBT test observations with SpaceX.
\end{IEEEbiographynophoto}

\vspace{-33pt}
\begin{IEEEbiographynophoto}{\ddueri} received the Ph.D. degree in Guidance and Control from the University of Washington in 2017. He is a Senior Manager of the Network Software Engineering division in SpaceX. 

\end{IEEEbiographynophoto}
\vspace{-33pt}
\begin{IEEEbiographynophoto}{\fschinzel} received the Ph.D. degree in Astrophysics at the Max Planck Institute for Radio Astronomy in 2011. He is an NRAO scientist and the lead of NRAO’s New Mexico Interference Protection Group (NM-IPG). He has served on many spectrum related committees and working group with many years of experience in spectrum policy and legal filing.
\end{IEEEbiographynophoto}

\vspace{-33pt}
\begin{IEEEbiographynophoto}{\miverson} received the M.S. degree in Computational and Applied Mathematics from the Colorado School of Mines in 2020. He is a Software Engineer in the Network Software Engineering team at SpaceX. He has been leading the development of the Telescope Boresight Avoidance (TBA) algorithm for the Starlink system. 
\end{IEEEbiographynophoto}

\vspace{-33pt}
\begin{IEEEbiographynophoto}{\jdonenfeld} received the B.S. degree in Mathematics and Computer Science from the Harvey Mudd College in 2021. He is a Software Engineer in the Network Software Engineering division at SpaceX. 
\end{IEEEbiographynophoto}

\vspace{-33pt}
\begin{IEEEbiographynophoto}{\bschepis} received the M.A. degree in Law and Philosophy from the University College London. He previously worked in the Satellite Policy division to lead the coordination efforts at SpaceX with NSF. 
\end{IEEEbiographynophoto}

\vspace{-33pt}
\begin{IEEEbiographynophoto}{\dgoldstein} received the Ph.D. degree in Aerospace Engineering from the University of Colorado Boulder in 2000. He is a Principal Guidance Navigation and Control Engineer at SpaceX. He has 30-plus years of Air Force and industry experience. 
\end{IEEEbiographynophoto}
\vfill

\end{document}